\documentclass{elsart3p}
\journal{Phys. Lett. B}
\usepackage{latexsym}
\usepackage{epsfig}

\usepackage{graphicx}
\usepackage{bm}
\usepackage{amsmath}
\usepackage{amssymb}
\newcommand{\be}{\begin{equation}}
\newcommand{\ee}{\end{equation}}
\newcommand{\bea}{\begin{eqnarray}}
\newcommand{\eea}{\end{eqnarray}}

\begin{document}

\begin{frontmatter}
\title{Three-form cosmology}
%\date{\today}
\author{Tomi S. Koivisto 
%\ead{t.koivisto@thphys.uni-heidelberg.de}
}
\and
\author{Nelson J. Nunes
%\ead{n.nunes@thphys.uni-heidelberg.de}
}
\address{Institute for Theoretical Physics, University of Heidelberg, 69120, Germany}

\date{\today}

\begin{abstract}

Cosmology of self-interacting three-forms is investigated. The minimally coupled canonical theory can naturally generate 
a variety of isotropic background dynamics, including scaling, possibly transient acceleration and phantom crossing.
An intuitive picture of the cosmological dynamics is presented employing an effective potential. Numerical solutions and
analytical approximations are provided for scenarios which are potentially important for inflation or dark energy.


\end{abstract}
\end{frontmatter}

%\keywords{Cosmology: Theory, Inflation, Dark Energy, N-Forms}
%\pacs{98.80.-k,98.80.Jk}

%\maketitle

\section{Introduction}
\label{introduction}

Phenomenology of cosmological scalar fields \cite{scalars} is studied extensively due to their conventional role in inflation and 
dark energy \cite{reviews}. Only recently the possibility of significance of higher spin fields has begun to attract some modest interest.
Vector inflation \cite{Ford:1989me}, using either time-like \cite{Koivisto:2008xf} or space-like \cite{Golovnev:2008cf} components, or featuring the vector as a 
curvaton \cite{v_curvaton} has been taken under consideration \cite{vectors} and the stability of the vector models has been analyzed in 
details \cite{himmetoglu}. 
Vector dark energy \cite{vde} could allow alleviation of the coincidence 
problem \cite{vde2} and even an electromagnetic origin of the cosmological term \cite{emde}. Spinor fields \cite{spinor} and Yang-Mills fields \cite{ym} have 
been explored also. 

The study of phenomenology of vector fields and their generalizations to forms is interesting since such fields appear in known and in speculative physics 
\cite{Eguchi:1980jx}. For example, the dynamics of Kalb-Ramond forms with dilaton couplings in string cosmology were studied in Ref.~\cite{kb}. Two-forms 
appear also in the framework of the asymmetric gravity \cite{tf1} as the antisymmetric contribution to the metric, and have been considered in cosmology 
\cite{tf2}. Forms, being intrinsically anisotropic, could be relevant for a dynamical 
origin of the four large dimensions \cite{ArmendarizPicon:2003qw}, modelling Lorentz violation \cite{lorentz} or the observed CMB anomalies \cite{anomal}. 
Form-driven inflation was suggested earlier this year \cite{forms,Koivisto:2009sd}. It was noticed that two-form inflation resembles much 
the vector inflation, having similar possibilities and problems, and that four-form becomes equivalent to scalar field or a higher order 
gravity theory, see though  \cite{Gupta:2009jy}. 

In the present Letter we focus on three-forms. 
It was proposed in Ref.~\cite{Koivisto:2009sd} that even the canonical field with 
a minimal coupling to gravity could naturally generate inflation, and here we show that such a three-form can also act as dark energy \footnote{Despite of this we 
might resist 
the temptation of calling the field ''trinflaton'' or ''trintessence''.}. 
The axion appears sporadically in cosmology, but with the assumption that it can be reduced to a (pseudo)scalar field. The fact that this is not always the case but 
fundamentally non-scalar fields can emerge at the low energy limit of string theory, has remained largely unexplored. The three-form however, sometimes 
allows a dual description as a scalar field. In the case of a non-quadratic
potential, the kinetic term of the scalar field is noncanonical. Such a model then becomes equivalent to k-inflation, and has been analysed before
\cite{threeform}. However, with a more complicated self-interaction or with any non-minimal coupling, the duality
with a scalar field breaks down. Thus, in specific cases the three-form cosmology reduces to a new perspective on effective scalar field cosmology, but
more generically it is a genuinely new model.

The structure of the Letter is as follows.
We will write down the basic equations in section \ref{Applications} and describe the features of the model in terms of an effective potential.
A convenient variable to describe is the comoving field $X(t)$, whose dynamics have some analogy with a scalar field. In section \ref{epochs} we discuss in more
detail some of the cosmological scenarios. It is shown that tracking solutions, where the energy density of the $X$ component scales like a
power of the scale factor, can be realized for a controllable number of e-folds. We also consider a kinetic phase of cosmology, the three-form domination
and the following epoch, which occurs in the interesting case that the three-form domination is transient. Finally we conclude in section \ref{conclusions}.

%%%%%%%%%%%%%%%%%%%%%%%%%%%%%%%%%%%%%%%%%%%%%%%%%%%%%%%%%%%%%%%%%%%%%%%%
\section{The model}
%%%%%%%%%%%%%%%%%%%%%%%%%%%%%%%%%%%%%%%%%%%%%%%%%%%%%%%%%%%%%%%%%%%%%%%%

\label{Applications}

We shall focus on a canonical theory minimally coupled to Einstein gravity. The action for a three-form $A_{\mu\nu\rho}$ can then be written as (omitting the indices)
\be \label{action}
S_A = \int d^4 x \, \sqrt{-g}\left(\frac{1}{2\kappa^2}R - \frac{1}{48}F^2 - V(A^2)\right) \,,
\ee
where $\kappa^2 = 8\pi G$. Here $F$ is the generalization of the Faraday form appearing in Maxwell theory,
\be
F_{\alpha\beta\gamma\delta} = 4 \nabla_{[\alpha}A_{\beta\gamma\delta]}.
\ee
We consider a flat FLRW cosmology described by the line element
\be
ds^2 = -dt^2+a^2(t)d{\bf x}^2.
\ee 
The nonzero components of the three-form are then given by
\be
A_{ijk}  =  a^3(t)\epsilon_{ijk}X(t). 
\ee
where we have considered, instead of the field $A$, the more convenient comoving field $X$. The relation between the squared invariant $A^2$ and the 
comoving field $X$ is then $A^2 = 6X^2$. We can thus consider the potential as a function of $X$, as we will do in the following. 

The equation of motion of the field $X$ is then:
\be
\label{eomX}
\ddot{X} = - 3 H \dot{X} - V_{,X} - 3 \dot{H} X  \,,
\ee
and a background perfect fluid evolves with
\be
\dot{\rho}_B = -3H \gamma \rho_B \,,
\ee
where $\gamma = -1 + p_B/\rho_B$,  
and these equations are subject to the Friedmann constraint
\begin{eqnarray}
H^2 &\equiv& \left(\frac{\dot{a}}{a}\right)^2 \nonumber \\
     &=& \frac{\kappa^2}{3} \left( \frac{1}{2} (\dot{X}+3H X)^2 + V(X)+ \rho_B \right) \,, \label{f1} \\
\dot{H} &=& -\frac{\kappa^2}{2} \left(V_{,X}X + \gamma \rho_B\right) \label{f2} \,.
\end{eqnarray}
We can thus define energy density and pressure of the field as
\begin{eqnarray}
\rho_X &=& \frac{1}{2} (\dot{X}+3H X)^2 + V(X) \,, \\
p_X &=& -\frac{1}{2} (\dot{X}+3H X)^2 - V(X) + V_{,X} X \,. 
\end{eqnarray}

\begin{figure}
\includegraphics[width=7.5cm]{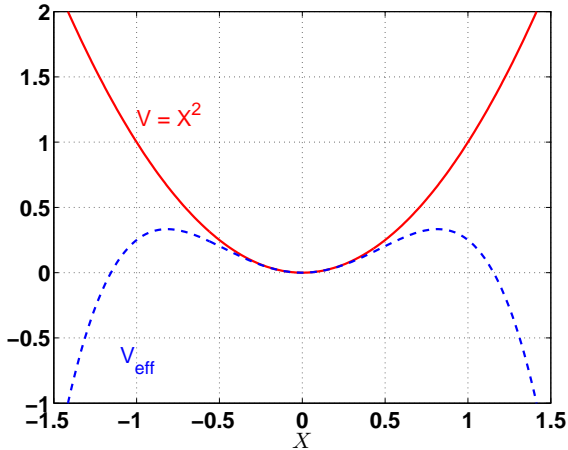}
\caption{\label{veff1} The potential (red, solid line) and the effective potential (blue, dashed line) for the quadratic potential. Units of $\kappa = 1$.}
\end{figure}

We can predict the evolution of the system by inspecting the equation
of motion of $X$, Eq.~(\ref{eomX}). Indeed, we can interpret the last two terms as the components of an effective potential with
\begin{eqnarray} \label{v_eff}
{V_{\rm eff}}_{,X} &=& V_{,X} + 3 \dot{H} X \nonumber \\
&=& V_{,X} \left(1- \frac{3}{2} X^2\right) - \frac{3}{2} \gamma \kappa^2 \rho_B X \,.
\end{eqnarray}
We illustrate in Figs.~\ref{veff1} the potential and the effective potential (up to a constant), when $\rho_B = 0$, for the simple power-law case. 

The equation of state parameter of the three-form, $w_X = p_X/\rho_X$,  can be written as
\be
\label{eos}
w_X = -1 + \frac{V_{,X}X}{\rho_X} \,.
\ee
It then follows that whenever the slope of $V(X)$ is negative (positive) if $X$ is positive (negative), the comoving field behaves as a phantom field. 
So the origin has some absolute meaning for this field, unlike in the case of a scalar. 
This brings us to a disclaimer of the intuitive picture we have just presented. Indeed, the validity of this picture is limited and the dynamics of 
%the field is more complicated. The reason is that the three-form couples to gravity in a different way than a scalar field; when writing 
\be
\ddot{X} + 3H\dot{X} + {V_{\rm eff}}_{,X} = 0
\ee
with $V_{\rm eff}$ given in Eq.~(\ref{v_eff}), this information is hidden in the friction term $H$ which now depends very differently on the field and its 
derivative than in the case of a scalar. We can see this by rewriting the Friedmann equation (\ref{f1}) as
\be
H^2 = \frac{1}{3} \frac{ V + \rho_B}{1-\kappa^2 (X'+3X)^2/6} \,,
\ee
where prime means differentiation with respect to $N = \ln a$.
In particular, we see that $X$ cannot be displaced further than $\kappa X = \pm\sqrt{2/3}$ for $X' = 0$, as this saturates the Friedmann equation. 

Despite the form of the effective potential for large values of $X$, this {\it does not} correspond to a runaway direction. To see this we must realise that $\kappa^2(X'+3X)^2 < 6$, from the Friedmann constraint and therefore 
$X' < \sqrt{6}/\kappa - 3X$ or $X'> -\sqrt{6}/\kappa - 3X$ for positive and negative $X$ respectively. In other words, when the initial position of $X$ takes large values, its 
velocity must necessarily be large and in the direction of decreasing absolute value of $X$. This makes it possible to consider scaling solutions in these 
models, as will be shown below.

%%%%%%%%%%%%%%%%%%%%%%%%%%%%%%%%%%%%%%%%%%%%%%%%%%%%%%%%%%%
\section{Cosmological epochs}
\label{epochs}

In this section we investigate the cosmological dynamics, focusing on a particular epoch at the time. We consider a scenario where there is a period of scaling
with the background scale factor, after which the field begins to approach the extremal point $\kappa X=\sqrt{2/3}$, but may not reach it but ends in the true minimum of the 
effective potential. We will see that the universe accelerates when the field is near the extremal point, this scenario could then describe the occurrence of a transient dark energy era after a scaling matter domination or, alternatively,  describe the early inflation of the universe. 
%Since the acceleration ends after the number e-folds given by the closeness of the extremal point, an exit from inflation is provided too. 

\subsection{Scaling and tracker solutions}
\label{scaling}

\begin{figure}
\includegraphics[width=7.5cm]{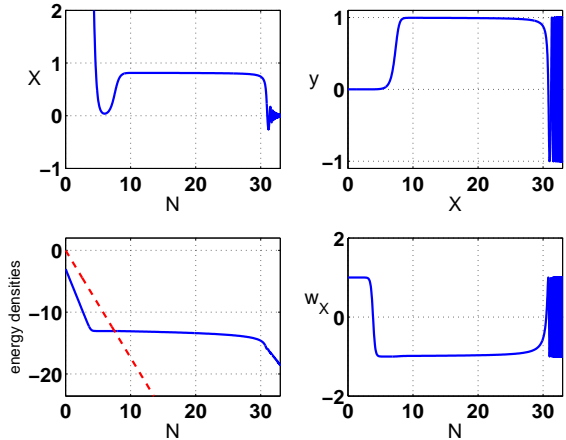}
\caption{\label{x0} 
Cosmological evolution as a function of the e-folding time $N=\ln{a}$ for the quadratic potential. In the upper left 
panel we see the field going through the unstable fixed point at $\kappa X=\sqrt{2/3}$ to oscillate around the stable minimum at $X=0$. The upper right panel shows 
the evolution of the combination $y = \kappa (X'+3X)/\sqrt{6}$. The lower left panel depicts the energy densities in a logarithmic scale; the dashed (red) line is matter and the 
solid (blue) line is the three-form. A brief tracking period is included in the figure, followed by two stages: the unstable fixed point and the 
oscillations around the stable minimum. The lower right panel shows the equation of state of $X$, exhibiting strong oscillations during the settling to the minimum. }
\end{figure}

We will now describe the evolution in a cosmological set up and comment on the possible applications in cosmological inflation and dark energy. In 
Fig.~\ref{x0}, we illustrate the evolution for the potential $V = X^2$. 
We can see that at early times the evolution of the energy density of the field $X$ is initially described by a tracking behaviour followed by a nearly constant value and at very late times it enters an oscillatory regime. We will 
now try to motivate this behaviour. 
We will start by taking the equation of motion when the background fluid is dominant, i.e., $\dot{H}/H^2 = -3\gamma/2$. Hence the equation of motion, written in terms of the time variable $N$, is approximately 
\be
(X'+3X)' -\frac{3}{2} \gamma (X'+3X) + \frac{V_{,X}}{H^2} = 0 \,.
\ee
Because we are interested in the evolution of the field from large values of $X$ we can for the cases of interest discard the contribution on the ratio of the derivative of the potential to the Hubble rate to obtain the simple solution 
\be
\label{genxsol}
X = A e^{3\gamma N/2} + B e^{-3N} \,.
\ee
Defining $y \equiv \kappa (X'+3X)/\sqrt{6}$ we see that $A = \sqrt{2/3}\, y_i/(1+\gamma/2)/\kappa$ and $B = X_i - A$. As we have seen before, large initial values of 
$X$ imply that $X' \approx -3X$ and equivalently, $y_i \ll 1$. Consequently, we can approximate
\bea
X = X_i e^{-3N} \,, \hspace{1cm} y = y_i e^{3\gamma N/2} \,.
\eea
The field $X$ scales as the background fluid provided 
$\rho_X \propto H^2 \propto e^{-3\gamma N}$. Given our system, this can only be possible if the kinetic term is negligible compared to the potential such 
that $V(X) \propto H^2$ and consequently, given the approximate evolution of $X$, the potential must be of power law type, $V(X) = V_0 X^n$ with $n = \gamma$. Let us then focus on power law potentials with  power arbitrary which in the case of $n \neq \gamma$ leads to a regime denominated ``tracker''. 

The scaling ($n= \gamma$) or tracking ($n \neq \gamma$) behaviour ceases when 
$3 H^2 y^2 \approx V(X)$. The number of e-folds elapsed is then found to be
\be
N_s = \frac{1}{3n} \ln \left( \frac{\kappa^2 V_i}{3 H^2_i y_i^2} \right) \,.
\ee
Hence we conclude that the number of e-folds of tracking increases for small values of $y_i$ and large ratios of $V_i/H_i^2$. We have tested these 
considerations numerically and found the numerical solutions behave as expected.

\subsection{Kinetic energy domination}
\label{kinetic}

The second epoch consists primarily by the fact that the field departs from the exponential behaviour and in some situations can even turn around in the potential and start climbing it. This is highly counter intuitive and deserves a closer look. Indeed the effect of the background fluid is of decelerating the field $X$. Keeping both contributions in Eq.~(\ref{genxsol}), we can compute the value of $N$ when the field turns around in the potential, 
\be
N_t = \frac{1}{3\left(1+\gamma/2\right)} \, \ln \left(\frac{2}{\gamma} \frac{B}{A}\right) \,,
\ee
and the value of the field at turn around
\be
X_t = A \left(\frac{2}{\gamma}\frac{B}{A}\right)^{\gamma/(\gamma+2)} + 
B \left(\frac{2}{\gamma}\frac{B}{A}\right)^{-2/(\gamma+2)} \,.
\ee
Of course this only happens if the background contribution is sufficiently large to completely stop the field. Nonetheless, the field now climbs up the potential and approaches $\kappa X = \pm\sqrt{2/3}$, with positive or negative velocity $X'$, respectively. From our discussion above, it should be clear now that the field cannot cross $\pm\sqrt{2/3}$ otherwise it would saturate the Friedmann equation, but instead asymptotically approaches this value until the background  contribution decreases sufficiently to allow for the field to once more roll downwards.

\subsection{Three-form domination}

At the next stage, the field is close to the extremal point and the background expansion then accelerates. 
To see this we introduce the appropriate
slow-roll parameter $\epsilon$ for this system 
\be
\label{epsdef}
\epsilon \equiv -\frac{\dot{H}}{H^2} = \frac{3}{2} \, 
\frac{V_{,X}}{V} \, X \, \left( 1- \frac{\kappa^2}{6}(X'+3X)^2 \right) \,.
\ee
We are now interested in investigating the inflationary dynamics near the 
extremum of the effective potential where $\kappa X = \pm\sqrt{2/3}$ and where the velocity of $X$ can be neglected, thus we find that $\epsilon$ can be well approximated by
\be
\label{epsapprox}
\epsilon \approx \frac{3}{2} \, 
\frac{V_{,X}}{V} \, X \, \left( 1- \frac{3}{2} (\kappa X)^2 \right) \,,
\ee
which allows us to immediately determine whether the universe is inflating for a given choice of the scalar potential at a given value of $X$. Using the original equation of motion for $X$ Eq.~(\ref{eomX}) and neglecting the $\ddot{X}$ contribution, it can be found that
\be
\label{dx1}
\kappa X' = -\frac{V_{,X}}{V} \left(1- \frac{3}{3} (\kappa X)^2 \right)^2 \,,
\ee
In general it is not possible to analytically solve for this equation given a choice of the potential, but fortunately it is for a pure power law potential, resulting that
\be
\kappa X = \pm \frac{\sqrt{2}}{3} \left(\frac{1+3 n N - 12 c}{n N - 4 c}\right)^{1/2} \,,
\ee
where $c$ can be fixed such that $X_i = X(N=0)$, and then $c = 1/(12-18 (\kappa X_i)^2)$.
For positive $n$ we can compute the value of $X_e$ at which inflation terminates, $\epsilon = 1$, giving
\be
\kappa X_e = \left(\frac{2}{3} - \frac{4}{4n}\right)^{1/2} \,,
\ee
Putting these results together we can compute the value of $X_i$ necessary in order to obtain $N_e$ e-folds of inflation,
\be
\kappa X_i = \pm \left(\frac{2}{3} - \frac{4}{9n} \, \frac{1}{1+2 N_e}\right)^{1/2} \,,
\ee
and clearly, the slow-roll condition on the velocity of the field (\ref{dx1}) must be satisfied. We observe that for larger values of $N_e$ then $\kappa X_i$ must be  closer to $\kappa X_i \approx \pm \sqrt{2/3}$. 

\subsection{Oscillations}

When the field reaches the minimum it enters an epoch of coherent decaying oscillations. We will now estimate the effective equation of state of the field during this period, $\langle w_X \rangle = -1 + \langle V_{,X} X/\rho_X\rangle$. We write 
$\rho \approx   V_{\rm max}$ as in scales much smaller than $H^{-1}$ it can be taken as a constant. We also approximate $\dot{X}^2 = 2(V_{\rm max} - V)$ as the extra term $3 H X \ll H \approx \sqrt{V_{\rm max}}$ can be neglected. Hence we have 
\bea
\langle w_X \rangle + 1 &=& \frac{\int V_{,X} X /\rho_X \, dt}{\int dt} \nonumber \\ 
&=& n \frac{ \int_0^{X_{\rm max}} V/V_{\rm max} \, \left(1-V/V_{\rm max}\right)^{-1/2} \, dX}{\int_0^{X_{\rm max}} \left(1-V/V_{\rm max}\right)^{-1/2} \, dX} \,.
\eea
For a power law potential it gives, surprisingly, the same result as for a canonical scalar field \cite{1983PhRvD..28.1243T},
\be
\langle w_X \rangle = \frac{n-2}{n+2} \,.
\ee
Thus for $n = 2$ the fields behaves as dust, $\langle w_X \rangle = 0$ and for $n = 4$ it mimics radiation, $\langle w_X \rangle = 1/3$.

\section{Discussion}
\label{conclusions}

We considered the evolution of the universe in the presence of three-forms. We assumed a canonical and minimally coupled action taking into account the possibility 
of self-interactions of the form field. Then a form with three differing spatial indices is compatible with an isotropic and homogeneous cosmological background. It 
turns out that such field, despite being canonical, quite generically violates the usual energy conditions. The strong energy condition is violated when 
\be \label{cond1}
V_{,X}X < \frac{4}{3}\left(\frac{1}{2}(\dot{X}+ 3HX)^2 + V\right ) \,\, 
\Rightarrow \,\, w_X < -\frac{1}{3}.
\ee
This happens quite easily. Slow roll is not required for accelerating behaviour, only that $V_{,X}X$ is not large compared to the energy density. The null energy 
condition is broken and the field behaves as phantom when
\be \label{cond2}
V_{,X}X<0 \,\,\Rightarrow \,\, w_X < -1 \,.
\ee
One notes that phantom divide crossing is possible for simple forms of the potential.
However, in analogy to the ghost pathologies appearing for several vector field models \cite{himmetoglu}, it turns out that 
in the phantom regime the three-form becomes rather generically unstable \cite{us}. 
In the null energy condition respecting regime, the three-form seems a very suitable culprit for the accelerating expansion which is 
believed to take place both at an early stage and at a late stage of the universe. 

We considered in detail a scenario where there is a period of scaling with the background scale factor, after which the field temporarily approaches the accelerating point 
$\kappa X= \pm \sqrt{2/3}$, but eventually terminates in the true minimum of the effective potential. Since the universe accelerates when the field is near the extremal point, 
this scenario could describe the occurrence of a transient dark energy era after a scaling matter domination. Escaping eternal de Sitter acceleration might help with the problem of 
S-matrix formulation in string theory \cite{Hellerman:2001yi}. Alternatively, this could describe the early inflation of the universe. 
Assuming the power-law potential, one may understand this 
as a reformulation of a massive k-essence model, where the kinetic dominance of the three-form corresponds to slowly rolling scalar and vice versa, however the
relation becomes untractable or ill-defined for slightly modified cases such as a double power law. Thorough clarification of these dualities, as well 
as quantitive computation of the observational predictions of these scenarios to generation and formation of structure will be presented elsewhere.

To conclude, we have shown that three-form cosmologies provide a well-motivated alternative to scalar field models. 

\section*{Acknowledgments}
N.J.N. is supported by Deutsche Forschungsgemeinschaft, TRR33.
This work was initiated at the workshop ''New horizons for modern cosmology'' at the Galileo Galilei Institute in Florence; we would like to thank the
institute for hospitality.

\end{document}